\documentclass[screen]{acmart}                   

\usepackage[utf8]{inputenc}        
\usepackage[nameinlink]{cleveref}  
\usepackage{dirtytalk}             
\usepackage{fontawesome5}          

\usepackage{pgfplots}
\usepackage{pgfplotstable}
\pgfplotsset{compat=1.17}

\newcommand{\fstar}{{\scriptsize\faStar}}

\copyrightyear{2026}
\acmYear{2026}
\setcopyright{cc}
\setcctype{by}
\acmConference[LAK 2026]{LAK26: 16th International Learning Analytics and Knowledge Conference}{April 27–May 01, 2026}{Bergen, Norway}
\acmBooktitle{LAK26: 16th International Learning Analytics and Knowledge Conference (LAK 2026), April 27–May 01, 2026, Bergen, Norway}
\acmDOI{10.1145/3785022.3785131}
\acmISBN{979-8-4007-2066-6/2026/04}

\begin{document}

\title{Fifteen Years of Learning Analytics Research: Topics, Trends, and Challenges}

\author{Valdemar Švábenský}
\authornote{Both authors contributed equally to this research and share the first author position.}
\orcid{0000-0001-8546-280X}
\affiliation{
    \department{Faculty of Informatics}
    \institution{Masaryk University}
    \city{Brno}
    \country{Czech Republic}
}
\email{valdemar@mail.muni.cz}

\author{Conrad Borchers}
\authornotemark[1]
\orcid{0000-0003-3437-8979}
\affiliation{
    \department{Human-Computer Interaction Institute, School of Computer Science}
    \institution{Carnegie Mellon University}
    \city{Pittsburgh, PA}
    \country{USA}
}
\email{cborcher@cs.cmu.edu}

\author{Elvin Fortuna}
\orcid{0009-0006-3344-6035}
\affiliation{
    \department{Department of Counseling, Educational Psychology \& Special Education}
    \institution{Michigan State University}
    \city{East Lansing, MI}
    \country{USA}
}
\email{fortun29@msu.edu}

\author{Elizabeth B. Cloude}
\orcid{0000-0002-7599-6768}
\affiliation{
    \department{Department of Counseling, Educational Psychology \& Special Education}
    \institution{Michigan State University}
    \city{East Lansing, MI}
    \country{USA}
}
\email{cloudeel@msu.edu}

\author{Dragan Gašević}
\orcid{0000-0001-9265-1908}
\affiliation{
    \department{Faculty of Information Technology}
    \institution{Monash University}
    \city{Melbourne}
    \country{Australia}
}
\email{dragan.gasevic@monash.edu}


\begin{abstract}
The learning analytics (LA) community has recently reached two important milestones: celebrating the 15th LAK conference and updating the 2011 definition of LA to reflect the 15 years of changes in the discipline. However, despite LA's growth, little is known about how research topics, funding, and collaboration, as well as the relationships among them, have developed within the community over time. This study addressed this gap by analyzing all 936 full and short papers published at LAK over a 15-year period using unsupervised machine learning, natural language processing, and network analytics. The analysis revealed a stable core of prolific authors alongside high turnover of newcomers, systematic links between funding sources and research directions, and six enduring topical centers that remain globally shared but vary in prominence across countries. These six topical centers, which encompass LA research, are: self-regulated learning, dashboards and theory, social learning, automated feedback, multimodal analytics, and outcome prediction. Our findings highlight key challenges for the future: widening participation, reducing dependency on a narrow set of funders, and ensuring that emerging research trajectories remain responsive to educational practice and societal needs.
\end{abstract}



\maketitle

\section{Introduction}

In 2011, the first LAK conference was held, and the following definition of learning analytics (LA) was established~\cite{Long-LAK11-LA-definition}:
\begin{quote}
\textit{Learning analytics is the measurement, collection, analysis and reporting of data about learners and their contexts, for purposes of understanding and optimising learning and the environments in which it occurs.}
\end{quote}

Throughout the next 15 years, LA has evolved from a small, emerging field~\cite{Siemens2013LA} into a mature field that diversified across learning contexts. The LA community has also grown, establishing itself through comprehensive books based on more than a decade of research. For example, the second edition of the Handbook of LA~\cite{Lang-handbook-la2022} reflects advances in research and practice, and a practical book on LA Methods and Tutorials~\cite{saqr-la-methods2024} helps researchers implement LA methods. Recently, the Society for Learning Analytics Research (SoLAR) has undergone a comprehensive process to converge towards an updated, joint definition of LA, aiming to better represent the changes in the field~\cite{SOLAR-LA-definition}:
\begin{quote}
\textit{Learning analytics is the collection, analysis, interpretation and communication of data about learners and their learning that provides theoretically relevant and actionable insights to enhance learning and teaching.}
\end{quote}

By shifting the focus to \say{actionable insights to enhance learning and teaching,} the new definition aims to address the criticisms that some prior LA research did not quite contribute towards improving learning. While foundational papers established the vision that LA is \say{about learning}~\cite{Gasevic2015LA}, looking back at the published work, recent surveys revealed that this was often not the case. For example, \citet{Motz2023LA} pointed out that publications at LAK and the Journal of LA (JLA) were often misaligned with the overarching goal of \say{closing the LA loop}~\cite{Clow2012cycle}, with over 70\% of papers not including a measure of learning. Another example is the LAK 2024 best paper by \citet{Kaliisa2024dashboardshype} on the lost promise of LA dashboards -- a popular topic at LAK -- showing there is no evidence that dashboards improve academic achievement. Critical examinations such as these demonstrate how important it is to assess, analyze, and reflect on our progress, which can help validate if the LA community stays on track with the objectives of the field as it develops.

\subsection{Goals and Scope of This Paper}
\label{subsec:intro-goals}

As the LA community now marks two important milestones simultaneously -- having celebrated 15 years of LAK and established the updated definition -- it is timely to assess the evolution of LA as a field. Specifically, we examined this by investigating the following research questions (RQs):

\textbf{RQ1:} \textit{Who has been involved in LAK research, where, and what changed over time?} Our first RQ was descriptive and practically oriented: to identify key players and global trends by tracking influential contributors.

\textbf{RQ2:} \textit{How does the funding landscape influence the research topics at LAK over time?} We examined relations between research topics and funding sources by treating topics as an evolving network. Because funders shape capacity and expertise, we analyzed whether specific funders correlate with accelerated development of topics or the formation of specialized clusters—viewing funding as a catalyst or constraint on the “economic development” of the LAK topic space.

\textbf{RQ3:} \textit{What are the topical trends in LA research globally and locally?} We inferred trends across research groups, recognizing LA’s branching into subfields~\cite{baker2021four} and its distinct “flavors” due to local practices, data, and privacy (e.g., divergent European initiatives~\cite{Nouri2019europe, Ferguson2015europe}). 

\subsection{Relevance for Learning Analytics Research and Practice}

To improve LA research and practice, we position our work in relation to LAK papers that aimed to understand the field from within. These include the above-mentioned study by \citet{Motz2023LA}, who investigated the impact of LA research through the lens of approaches to validation and evaluation, and the survey by \citet{Kaliisa2024dashboardshype} who examined the dashboard hype. Unlike review papers that involve manual coding, we employed unsupervised machine learning (ML)~\cite{Pedregosa-scikit-learn} and natural language processing (NLP)~\cite{nltk} to analyze the development of LA. This process was inspired by how data science disciplines study the rise and fall of trends and innovation~\cite{hidalgo2007product, Tierney2020creative}. Moreover, this represents an opportunity to shed light on how our field developed, concluding the first 15 years of LAK as it begins a new chapter.

\subsection{Contributions}

Building on articles that critically examined LA~\cite{Kaliisa2024dashboardshype, Ferguson2025doctoral}, we analyze trends across all 936 papers published at LAK over its 15 years. Our study provides methods and insights to guide more rigorous research and practice, enabling the community to enhance LA applications and, ultimately, improve human learning. We make the following contributions:
\begin{itemize}
    \item We describe the evolution of the LA landscape across all 15 years of LAK, from its inception to the latest trends. Our focus is on mapping LA topics and communities worldwide, providing a practical and global perspective.
    \item We identify the key authors and countries focusing on LA, as well as funding agencies supporting LA research. This information can help both newcomers entering the field, such as Ph.D. students and postdocs looking for job positions, as well as established researchers and practitioners, to navigate the state of the art.
    \item Our analysis reveals the interconnection between funding providers and specific research topics, showing that funders do not choose topics randomly; instead, certain topics are more likely to be funded.
    \item Beyond the descriptive insights, as in most review papers, we apply inferential statistics, ML, and NLP methods to discover overarching trends and propose how the field can move forward.
    \item We publicly provide our curated research dataset, which describes 936 full and short papers published at LAK 2011--2025, as well as the associated analytical code~\cite{supplementary-materials}. Other researchers can adopt or adapt it to validate our results or extend our analytics by focusing on other areas not covered by this paper.
\end{itemize}

\section{Related Work}
\label{sec:related-work}

Although LAK has been historically distinct from other communities~\cite{baker2021four}, such as Educational Data Mining (EDM), Artificial Intelligence in Education (AIED), and Learning at Scale (L$@$S), it has some overlap with them~\cite{Bittencourt2024paied}. This section reviews surveys across LAK and the related communities. 

Early work by \citet{Dawson2014current} analyzed the first three years of LAK and papers from selected journals, focusing on two aspects: citation networks and authors. The most cited papers were largely conceptual, calling for more empirical work. Authors were grouped into many \say{small cliques with few highly interconnected authors.} The survey encouraged more interdisciplinary teams and envisioned that new researchers would undertake their Ph.D. or postdoc in LA. Extending this line of work, \citet{Waheed2018bibliometric} analyzed 17 years of LA publications and identified the United States (USA), Spain, Australia, the United Kingdom (UK), and Germany as the top contributing countries. They identified \say{smaller research communities and a lack of collaboration among such clusters,} aligning with the previous survey. The survey highlighted the need for future research to examine a broader range of topics. \citet{Du2020edmreview} added to this perspective by analyzing 1,219 EDM studies (64\% empirical studies, 19\% proof of concept, and the rest were frameworks or reviews) across four major areas: performance prediction, decision support, behavior detection, and learner modeling. 

Building on this foundation, recent cross-venue syntheses converge to situate LAK’s topical emphases relative to EDM, its position within cross-community keyword and collaboration networks, the doctoral-consortia pipeline into the field, and the emerging research frontiers likely to shape the next phase. \citet{Chen2020lakedm} compared all LAK and EDM papers to date, noting 11 groups of research topics. Despite notable overlap between LAK and EDM, \textit{Engagement Patterns \& Resource Use} was more commonly researched at LAK, whereas \textit{Predictive \& Descriptive Analytics} were more common at EDM. Among the prominent regions are Europe and Oceania for LAK, and the Americas and East Asia for EDM, with both conferences having many new authors every year. Extending this, \citet{Zheng2022aied} analyzed eight years of AIED, EDM, LAK, and L$@$S. They found higher education, visualizations, and self-regulated learning (SRL) as unique keywords for LAK. Also, LAK was considered \say{the most connected and the most centralized community,} fostering greater participation also outside of the USA (the most contributing country overall). However, they warned that using \say{synonymous but distinct terminologies to describe the same concepts in different cliques may hinder inter-group communication.} \citet{Ferguson2025doctoral} showed that many newcomers attending the LAK doctoral consortium return to LAK in the future, publish more papers, and even take on organizational roles. Finally, \citet{feng2025aiedheadedkeytopics} explored key topics in the broad AIED domain, including papers at LAK 2020--2024. They found that intelligent tutoring systems, NLP, and MOOCs continue to play important roles, with future frontiers being large language models (LLMs), generative AI, human-AI collaboration, and multimodal LA.

Our paper is the first to holistically analyze all 15 years of LAK. Although other scientific fields have used methods similar to ours to study trends and understand their respective disciplines~\cite{Chen2024HCIpapers, feng2025aiedheadedkeytopics, Lunn2021sigcse}, such a paper was missing at LAK.

\section{Dataset of LAK 2011--2025 Papers}

\begin{figure*}[!ht]
\centering
\begin{tikzpicture}
    \begin{axis}[
        width=15cm,
        height=4.5cm,
        xlabel={Year},
        ylabel={Paper count},
        xmin=2010, xmax=2026,
        ymin=0, ymax=125,
        xtick={2011,2012,2013,2014,2015,2016,2017,2018,2019,2020,2021,2022,2023,2024,2025},
        ytick={0,25,50,75,100,125},
        grid=both,
        major grid style={line width=.2pt,draw=gray!50},
        minor grid style={line width=.1pt,draw=gray!20},
        xticklabel style={/pgf/number format/assume math mode=false, /pgf/number format/1000 sep={} },
        nodes near coords,
        nodes near coords style={font=\small, anchor=south},
        legend style={at={(0.08,0.95)}, font=\footnotesize, anchor=north},
    ]

    \addplot[
        color=blue,
        mark=*,
        line width=1pt
    ] coordinates {
        (2011,24)
        (2012,40)
        (2013,40)
        (2014,38)
        (2015,59)
        (2016,62)
        (2017,64)
        (2018,60)
        (2019,69)
        (2020,80)
        (2021,71)
        (2022,62)
        (2023,71)
        (2024,95)
        (2025,101)
    };
    \addlegendentry{Papers}
    
    \addplot[
        blue,
        dashed,
        domain=2011:2025,
        samples=2
    ] {62.4};
    \addlegendentry{Average}

    \addplot[
        red,
        dashed,
        domain=2011:2025,
        nodes near coords={}, 
    ] {4.275*x - 8564.55};
    \addlegendentry{Trend}
    
    \end{axis}
\end{tikzpicture}
\caption{Evolution of the number of published papers across all 15 years of LAK (936 papers total; 62.4 papers per year on average).}
\Description{This graph is a line chart showing individual years 2011--2025 on the x-axis, and the number of published papers on the y-axis. It also shows an average and an overall trend line, which is increasing.}
\label{figure:paper-count}
\end{figure*}
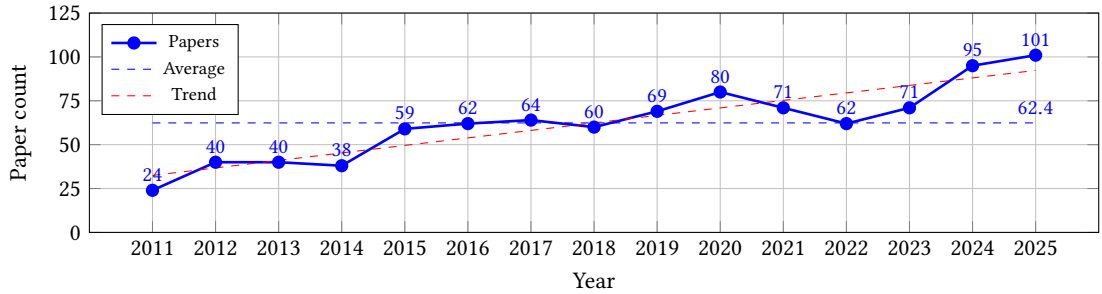

Our research dataset comprises the metadata and content of all LAK papers published over the 15 years of LAK, from 2011 to 2025, by combining two separate datasets. We focused only on publications in the main track, i.e., full and short research papers. Therefore, we omitted submissions from the companion proceedings, such as posters and demos. These were not as stringently peer-reviewed as the main track, and the technical depth of those submissions varies. As shown in \Cref{figure:paper-count}, there were 936 full and short papers in total. This count was manually established by the authors, since the front matter of the proceedings sometimes contained an off-by-one error regarding the number of accepted full/short papers. For each LAK paper, we collected the data described in \Cref{table:dataset_fields} relevant to our RQs. 

\begin{table*}[!ht]
\centering
\caption{Each LAK paper (row in the dataset) is characterized by the following fields (columns in the dataset). Columns prefixed by the \fstar\ symbol are available for all papers (2011--2025). The remaining columns were captured only in our custom dataset (2020--2025).}
\scriptsize
\begin{tabular}{|l|l|l|l|}
\hline
\textbf{Column name} & \textbf{Description / meaning} & \textbf{Possible number of entries per paper (per row)} & \textbf{Value constraints} \\ \hline

\fstar\ \textit{DOI} 
 & Unique identifier of a paper, e.g., 10.1145/1234567.1234567
 & Always 1
 & Numeric-like string
 \\ \hline

\fstar\ \textit{Year} 
 & Year of the paper publication in the LAK proceedings
 & Always 1
 & Numeric, from 2011 to 2025
 \\ \hline
 
\fstar\ \textit{Title} 
 & Author-provided title of the paper
 & Always 1
 & Short string
 \\ \hline
 
\fstar\ \textit{Author} 
 & Name(s) of the paper author(s)
 & From 1 to 15 (observed maximum)
 & Short string
 \\ \hline
 
\fstar\ \textit{Affiliation} 
 & Institution for each author
 & From 1 to 15 (depending on the \textit{author} field)
 & Short string
 \\ \hline
 
\fstar\ \textit{Country} 
 & Country for each affiliation
 & From 1 to 15 (depending on the \textit{affiliation} field)
 & Short string
 \\ \hline
 
\fstar\ \textit{Abstract} 
 & Author-provided abstract of the paper
 & Usually 1, rarely 0 (some papers do not list abstract)
 & Long string (plain text)
 \\ \hline
 
\textit{Keywords} 
 & Author-provided keywords for the paper
 & From 0 to 10 (observed maximum)
 & Short string
 \\ \hline

\textit{Article type} 
 & Full paper (called \say{Research Article} by ACM) or short paper
 & Always 1 (for 2020--2025 papers)
 & \say{Research} or \say{Short}
 \\ \hline
 
\textit{Acknowledgments} 
 & Author-provided acknowledgments
 & Often 1, sometimes 0 (some papers do not list acks)
 & Long string (plain text)
 \\ \hline
 
\textit{Funding} 
 & Funding source inferred from the acknowledgments
 & From 0 to 11 (observed maximum)
 & Short string
 \\ \hline

\end{tabular}

\label{table:dataset_fields}
\end{table*}

\subsection{LAK 2011--2019: Extension of the Ochoa Dataset}
\label{subsec:data:ochoa}

For the first nine years of LAK, we started with the public dataset by Ochoa~\cite{Ochoa-dataset}, which contains most of the fields needed for our study (see \Cref{table:dataset_fields}). To preserve only full and short papers in the dataset, we removed all entries that were 1 or 2 pages long. The shortest published paper had 3 pages~\cite{Siemens2012learning} (among the recent papers, the shortest paper had 4 pages~\cite{Berland2023joint}). Then, we manually removed the remaining submissions outside the main track, some of which were 3 pages or longer. We double-checked that for each year, the paper count corresponds to that established in \Cref{figure:paper-count}.

Since the Ochoa dataset lacks authors’ affiliations, we developed custom code to add affiliations and corresponding countries as extra columns. For dual-country cases (48 out of 1333 affiliations, 3.6\%), we kept the first affiliation. We added \texttt{Unknown} for affiliations (0.2\%) of two independent researchers without a listed country. To resolve inconsistent encoding for the region of Hong Kong (e.g., HK, Hong Kong SAR China, and CN, China), we standardized all entries to the former. Finally, we manually verified all affiliations and country codes and corrected a small number of discrepancies.

\subsection{LAK 2020--2025: Our Own Data Collection}
\label{subsec:data:our}

For LAK 2020--2025 papers, we created the dataset from scratch. For each year, we downloaded the entire proceedings from the ACM Digital Library as a PDF. Our custom Python script was used to parse the proceedings, scrape relevant information from the PDFs, and store it in a table. Based on DOIs in this table, the script then checked CrossRef and DBLP databases for the remaining metadata and merged them with our entries. The structure of the resulting file is a superset of the Ochoa dataset structure, i.e., it includes all the data types and additional information (keywords, article type, acknowledgments, funding -- see \Cref{table:dataset_fields}). After creating the dataset, we performed several rounds of manual checks and validation post-processing, such as standardizing countries in affiliations, converting keywords to lowercase, and removing extraneous characters. Our dataset and code are publicly available as supplementary materials~\cite{supplementary-materials}.

\section{RQ1: Who Has Been Involved in LAK Research, Where, and What Changed Over Time?}
\label{sec:rq1}

\subsection{Authors at LAK}
\label{subsec:rq1-authors}

To obtain insights about authors, we used the \texttt{nameparser} module in Python~\cite{Gulbranson-nameparser} to perform unified parsing of names in our dataset. Each name was normalized, with middle names/initials removed, as well as punctuation (except for hyphens and apostrophes, which are commonly used in names). Then, we verified the intermediate dataset and manually corrected discrepancies, such as rare typos. Other than that, we used the names as listed by the authors themselves.

\subsubsection{Repeat Authors, Newcomers, and Former Participants}

Across all 936 LAK papers, there were 2,033 unique authors. Out of these, 1,440 (70.8\%) published only a single paper. As for the repeat authors, 304 (15.0\%) published two papers, 122 (6.0\%) published three papers, 125 (6.1\%) published between four and nine papers, 33 (1.6\%) published between 10 and 19 papers, and the remaining nine (0.4\%) published 20 or more papers. Almost all of these prolific authors started publishing at LAK in its initial years (2011--2015), and almost all of them continued to publish until 2025, representing stable contributors who shaped the field. Two of these authors were also identified as the top authors in a 2018 survey of a broader LA landscape~\cite{Waheed2018bibliometric}, and six were among the top ten most central according to~\citet{Zheng2022aied}.

At the same time, LAK welcomes many newcomers. In the past three years (from 2023 to 2025), 574 new authors appeared, representing 28.2\% of all 2,033 authors. Out of these new authors, 105 have already become repeat authors. To provide context, we analyzed the overall author retention rate, looking at the probability of whether an author published at LAK again within 3 or 5 years after publishing their first LAK paper. Within a 3-year window, the rate fluctuated between 15 and 30\%. Increasing the window to 5 years also increased the rate to 20--45\% (in both cases except LAK 2011, where the rate was much higher). The rate oscillates across time windows per year, not showing any upward or downward trend in recent years, meaning that the consistent \say{membership} in the LAK community is a challenge.

From another perspective, 966 (47.5\%) of all authors have not published in the past five years (i.e., their last paper was in 2020 or earlier), and 317 (15.6\%) have not published in the past ten years, representing a dropout. Overall, the LAK community is actively evolving: it has a stable base of a small group of prolific authors, a solid newcomer rate (however, with fluctuating retention rates), and 70\% one-time contributors who have not (yet) published again.

\subsubsection{Collaborations}

Out of all 936 papers, 39 (4.2\%) have a sole author, 391 (41.8\%) have two or three authors, 313 (33.4\%) have four or five authors, 132 (14.1\%) have six or seven authors, and the remaining 61 (6.5\%) have eight or more authors (maximum 15). The median number of authors is four, with an average of 3.87. Overall, we observed an increase in the number of authors per paper in recent years. This increase is notable, especially compared to an early LAK survey by \citet{Dawson2014current}, where most papers were identified as having 2--3 authors.

Looking at collaborators who co-authored two or more papers together, there are 913 unique pairs of collaborators. Out of these, 600 pairs (65.7\%) wrote two papers together, 196 pairs (21.5\%) wrote three papers together, 98 pairs (10.7\%) wrote between four and nine papers together, and the remaining 19 pairs (2.1\%) wrote ten or more papers together. The pairs with the most collaborations also belong among the most prolific authors. Overall, a substantial number of stable \say{cliques} repeatedly publish together -- a continuing trend observed as early as in LAK 2014~\cite{Dawson2014current}.

\subsection{Countries at LAK}
\label{subsec:rq1-countries}

When analyzing global contributions and cross-country collaboration, we considered only the unique countries per paper; that is, if a paper had multiple authors from the same country, it was counted only once. This method avoids inflating the importance of a repeated occurrence within a paper with multiple authors.

\subsubsection{LAK Presence}

Across all 936 LAK papers, there were 50 unique countries/regions (excluding \say{Unknown}), of which 48 are United Nations member states. Out of the 193 United Nations members, this is a 24.9\% coverage. The top ten countries/regions are, with their paper count: USA (451), Australia (191), UK (111), Canada (67), Germany (61), Netherlands (49), Spain (27), Hong Kong (26), China (26), Brazil (23). Among these, the top five have been publishing at LAK since its inception in 2011. These findings highlight the majority presence of the Global North countries. There was only one new country in the past three years -- Costa Rica. The widening of the range to the past five years (2021--2025) reveals four more newcomers: Croatia, Saudi Arabia, Iran, and Thailand. On the contrary, 11 countries/regions have not published at LAK in the past five years, representing a dropout. Improving the retention of the broader international community could be one of LAK's future goals; we expand on this topic with actionable suggestions in \Cref{subsec:implications}.

\subsubsection{Collaborations}

Most papers, 653 (69.8\%), were authored by a single country. Next are 210 papers (22.4\%) with two different countries, and 55 papers (5.9\%) with three countries. The median number of countries is one, and the average is 1.38. Interestingly, the cross-country collaboration is often across continents. For example, the USA with Australia, the UK, or China; and Australia with the UK, Brazil, or Canada; mirroring the most prominent countries. 

Speaking to the field's maturity and increasing collaboration, we find a significant increase in the average number of represented countries per paper over the years ($r = 0.62, p = 0.015$). This metric counts the unique countries of all co-authors per paper and averages this count by year. However, unlike the increasing number of authors, the trend in increasing cross-country collaboration is less strong and slightly fluctuates. Fostering even more international collaboration could reveal interesting insights across different educational contexts; we revisit this topic in \Cref{subsec:implications}. 

\section{RQ2: How Does the Funding Landscape Influence the Research Topics at LAK Over Time?}
\label{sec:rq2}

\subsection{Methods}

To identify topical trends, we used two approaches detailed below: keyword-based and representation learning-based. We then developed a topic-funder network from the topic and funding columns to analyze how these relationships evolved between 2020 and 2025 (see \Cref{subsec:data:our} for the rationale behind this range).

\subsubsection{Keyword-Based Approach}

We standardized the topic and funder labels to correct for typos and label variations. Similarity between topic strings was measured by Levenshtein (edit) distance, which calculates the minimum number of single-character edits needed to change one string into another. We then applied a fuzzy threshold to identify potential typos. A topic was considered a typo if its Levenshtein distance from a \say{canonical} (correct) string was less than 10\% of the canonical length. Next, we compared each label against our initial list of canonical strings. If a string's Levenshtein distance fell below the fuzzy threshold, it was assigned to the closest matching canonical string. If a topic had no close matches, it was added as a new canonical topic, which allowed our list to grow and adapt to the data, and vice versa for funder labels. Papers were removed if they did not list a funder or if the funder was a country (e.g., only \say{UK}).

To identify communities of related research topics, we utilized Louvain clustering from the \texttt{igraph}~\cite{csardi2006igraph} package in R (version 4.4.2)~\cite{R}. This algorithm detects communities in large networks~\cite{zhang2021improved}: nodes are grouped into clusters that are more densely connected to each other than to nodes in other parts of the network. These communities represent groups of topics frequently appearing together, reflecting a shared research area or theme.

\subsubsection{Representation Learning Approach}
\label{subsec:rq2-representation-learning}

Representation learning automatically maps high-dimensional raw data (such as text) into a lower-dimensional vector space that preserves semantics. We applied it to the abstracts of LAK papers, as abstracts provide concise summaries of the research focus and allow us to capture topical similarities across the corpus. Compared to keyword matching, representation learning may capture more nuanced semantic relations beyond exact word overlap. It facilitates scalable analysis across hundreds of papers, allowing us to quantify how similar papers are over time, how topics evolve, and which funders or countries contribute to different areas of the field.

We used the Python \texttt{SentenceTransformer} library (\texttt{all-MiniLM-L6-v2} model) to obtain dense vector embeddings for each abstract. This model is used for semantic similarity tasks and provides a balance of accuracy and efficiency (compared to larger embeddings, such as LLMs). Each abstract was thus represented as a 384-dimensional vector.

To identify topical groupings, we applied $k$-means clustering to the embeddings. The number of clusters was determined using the silhouette method, which evaluates the quality of clustering solutions by measuring the similarity of each document to its assigned cluster compared to other clusters. This prevents overfitting to too many fine-grained clusters. Across candidate values of $k$, six clusters yielded a good balance between internal coherence and interpretability. 

Clusters were validated in two steps. First, two independent coders manually interpreted the six clusters by reviewing the abstracts representative of each cluster. To facilitate this process, we computed cosine similarities between each abstract and the average cluster centroids, and selected the three most prototypical abstracts per cluster. This ensured that coders worked with the most characteristic texts for each group. Second, the author team—ranging from 5 to 15 years of experience in LA—reviewed and refined the interpretations. This iterative process provided both inter-rater validation and expert triangulation, ensuring the clusters were meaningful, robust, and aligned with domain knowledge.

\subsubsection{Bipartite Network}
\label{subsubsec:rq2-bipartite}

We calculated a bipartite network~\cite{liew2023methodology, kevork2022bipartite} to analyze relations between research topics and funding sources. A bipartite network consists of nodes divided into two distinct sets, \textit{X} and \textit{Y} (topics and funders in our case), and only connections between two nodes in different sets are allowed. 

Two separate topic-funder spaces were created, where nodes represented funders with either (1) topic communities (identified via keywords) or (2) topic clusters (identified via representation learning), to build a network analyzing how these relationships evolved. In this model, a topic's influence is measured by its connections to a diverse range of funders, and a funder's impact is seen in its ability to support and shape multiple topics. Various internal seed funds from individual universities were grouped under the label \say{institutional support funds} (ISF).

In our bipartite network, the edges represent a funding-topic relationship. A link exists between a topic and a funder if that funder has supported research related to that topic. We analyzed network connectivity using node and edge counts (number of topic-funder connections), density (overall interconnectedness), average degree (number of connections per node), and modularity, which indicates how well the network can be divided into distinct communities (higher values = topics and funders are grouped into tight-knit clusters; lower values = more interconnected, less fragmented clusters).

\subsection{Results}
\label{subsec:rq2-results}

\subsubsection{Topical Centers}

\begin{figure*}[!ht]
\centering
\includegraphics[width=\linewidth]{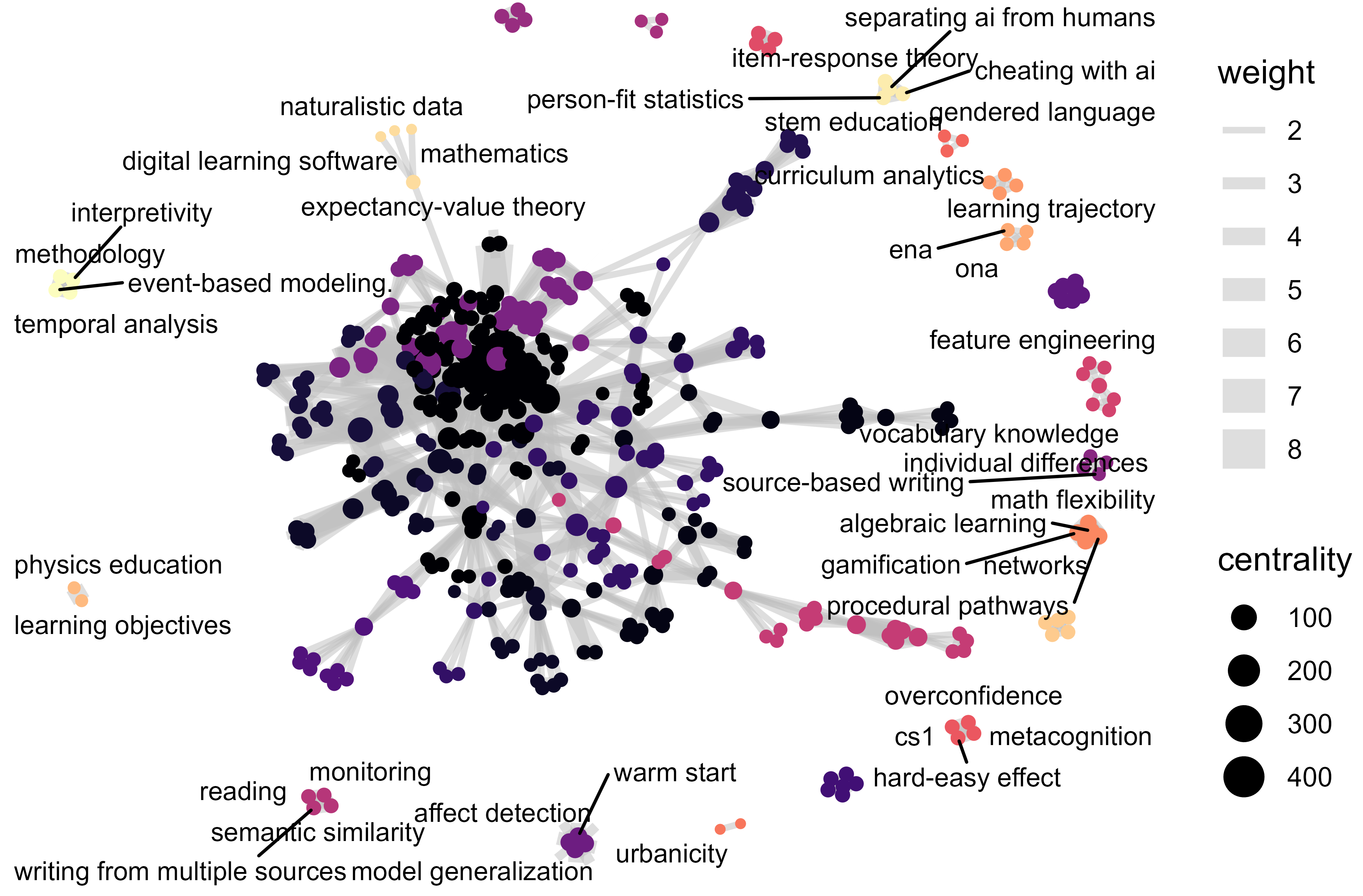}
\caption{Relations between topic communities (Louvain clustering) in one-mode projection of the bipartite network. Abbreviations: cs1 = introductory computer science, ena = epistemic network analysis, one = ordered network analysis, ai = artificial intelligence.}
\Description{This figure shows a network of related topic communities in the form of undirected graphs, with a core center and peripheral communities around it.}
\label{fig:topic_communities}
\end{figure*}

A one-mode projection was used to analyze the links between keyword topics, and Louvain clustering was applied to create topic communities. This network, where nodes represent the keyword topics, has a weighted edge between two topics if they co-occurred in at least one paper. The weight of the edge corresponds to the frequency of co-occurrence, and the color represents topic communities. In total, 29 topic communities were identified. Community~1 was the largest (black in \Cref{fig:topic_communities}), with 111 topics including association rule mining, clustering, sequence analysis, multimodal data, and student engagement. \Cref{fig:topic_communities} also shows that community~1 is well-connected to others. The second-largest community (purple in \Cref{fig:topic_communities}) consisted of 63 topics, including teaching analytics, collaborative problem-solving, computer-supported collaborative work, social networks, reflection, and data storytelling. Smaller communities may suggest highly specialized or niche areas of LAK research. The smallest communities had only two topics each, for example, cognitive evaluation theory and analytics-driven game design elements.

To reduce the number of topic nodes, we identified six topical centers by clustering paper abstracts (see \Cref{subsec:rq2-representation-learning}) and then manually interpreting each cluster. Below, we summarize each center with representative examples from prototypical papers (with their year of publication in parentheses) surfaced via centroid similarity.

\textbf{SRL \& Study Strategies} covers how learners regulate their learning and how process traces reveal learning tactics and strategies. Examples include micro-level SRL processes within distinct tactics detected from clickstream data (2021), the role of positive/negative self-regulated strategies in shaping online activity and achievement (2016), and process-oriented \say{observed study strategies} that predict performance and support early feedback (2020).

\textbf{Dashboards \& Theory} connects learning dashboards to learning science theory and the broader learning analytics cycle. Representative work includes a systematic review of learner-facing dashboards and their theoretical grounding (2018), early stakeholder perspectives on the scope and expectations for LA (2012), and an influential cycle that emphasizes \say{closing the loop} with timely feedback to learners and teachers (2012).

\textbf{Social Learning} draws on forum/network methods and their relationship to learning. Examples include social processes in persistent MOOC cohorts (2016), complementary constructions of interpersonal (reply) versus intergroup (co-participation) positioning (2020), and controls showing that some network indicators largely reflect posting activity while clustering captures richer social structure (2020).

\textbf{Automated Feedback} addresses systems that generate formative feedback on open responses. Illustrative work spans LLM-based essay scoring and feedback that supports human–AI co-grading (2025), a domain-agnostic essay feedback tool linked to improved outcomes (2015), and automated grading of open-ended mathematics responses with analysis of data scale effects (2020).

\textbf{Multimodal Analytics} focuses on rich, sensor-based data for collaborative and embodied learning. Examples include a group-work platform (2024), in-the-wild analytics of healthcare simulations with social, epistemic, and physical traces (2017), and a low-cost, open, easy-to-deploy device for varied classroom contexts (2024).

\textbf{Outcome Prediction} targets early and accurate prediction of success/risk at course or assessment scales. Representative studies include collaborative multi-regression models that share structure across students while personalizing predictions (2015), using the first 2–5 hours of logs to anticipate end-of-year external assessments across products and countries (2025), and engagement-aware knowledge tracing for targeted interventions (2019).

\subsubsection{Funding by Topic Center}

\begin{figure*}[t]
\centering
\includegraphics[width=\linewidth]{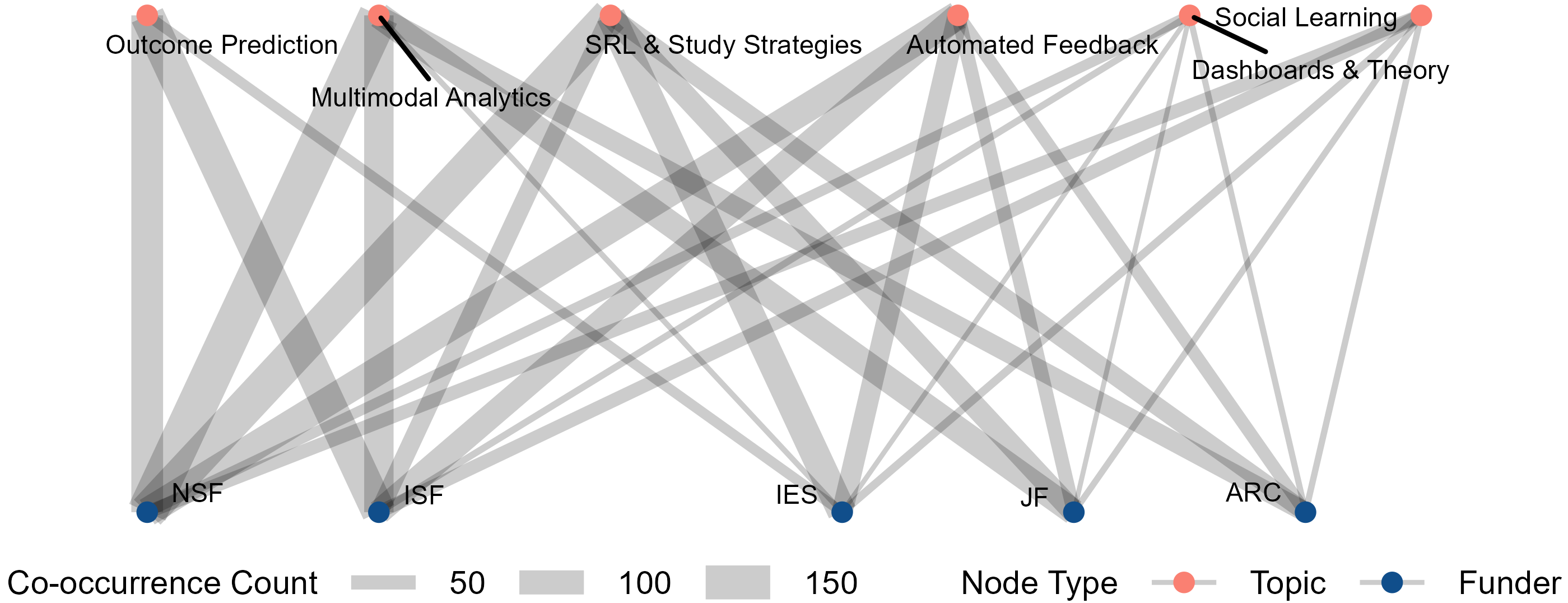}
\caption{Bipartite network between topic centers (abstracts) and top 5 funders of LAK from 2020-25. NSF = National Science Foundation, ISF = Institutional support funds, IES = Institute of Education Sciences, JF = Jacobs Foundation, ARC = Australian Research Council.}
\Description{A bipartite graph showing six nodes with topic centers on one side, and five top funders on the other side, with stronger interconnections between certain pairs.}
\label{fig:bipartite}
\end{figure*}

\begin{figure*}[t]
\centering
\includegraphics[width=\linewidth,trim=0 0 0 6,clip]{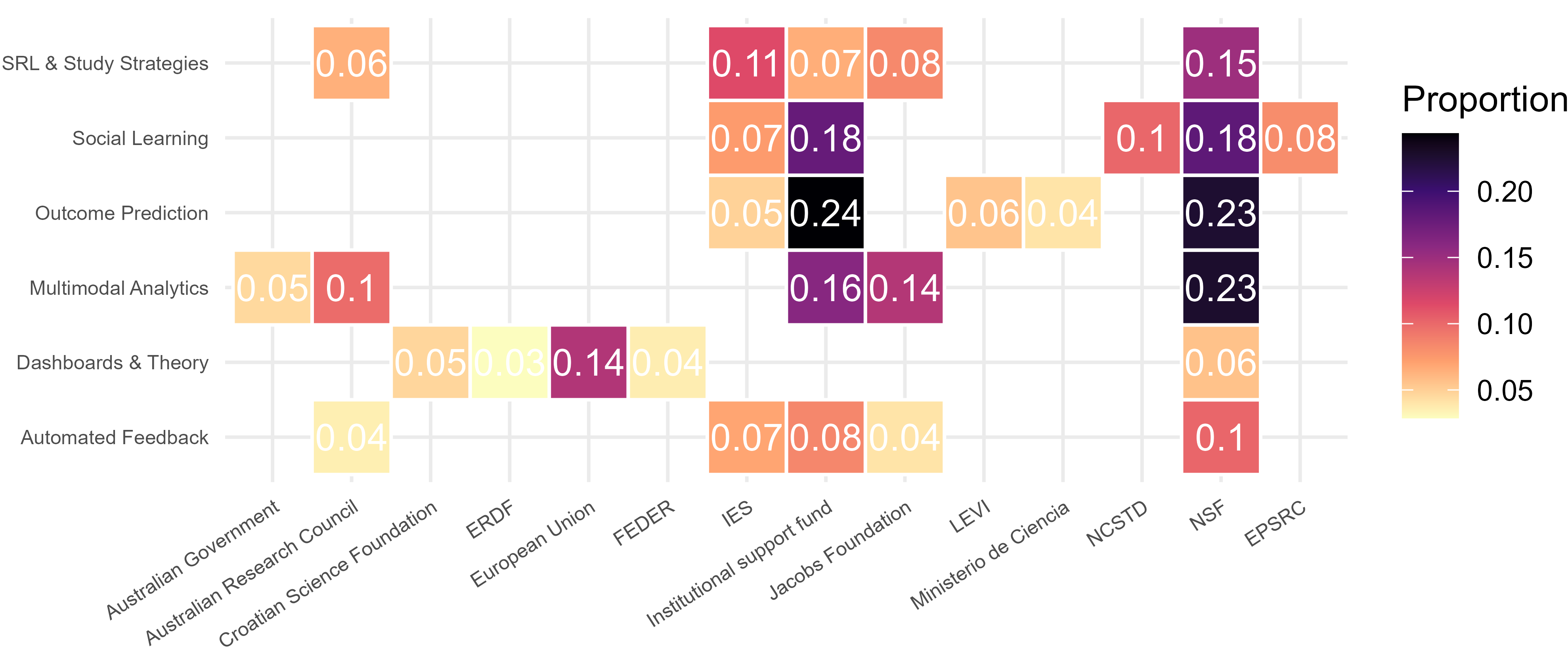}
\caption{Proportion of top funders (x-axis) who support topic centers (y-axis). NCSTD = National Council for Scientific and Technological Development, EPSRC = UK Engineering and Physical Sciences Research Council, IES = Institute of Education Sciences, LEVI = Learning Engineering Virtual Institute, NSF = National Science Foundation, ERDF = European Regional Development Fund.}
\Description{A heatmap showing that certain funders support some of the six topic centers more prominently.}
\label{fig:topic_centers}
\end{figure*}

The number of unique funders in each of the six years (2020--2025) is 46, 46, 26, 29, 61, and 55, respectively. The bipartite network revealed that the National Science Foundation (NSF) and institutional support funds (ISF, see \Cref{subsubsec:rq2-bipartite}) supported a diverse portfolio of projects related to different topic centers. NSF largely funded research on outcome prediction, multimodal analytics, SRL and study strategies, and automated feedback. ISF funded similar topics, but with more funds going to social learning compared to NSF (see \Cref{fig:bipartite}).

We assessed funder-topic links via the distribution of funding across topic centers. We calculated the proportion of funded versus unfunded topic centers and used a Chi-square test of independence to check if there were significant links between funders and the observed number of funded topic centers. Results showed a significant relationship, $\chi^2 = 2926$, $df = 60$, $p < 0.0001$, indicating that certain funders targeted central topics at rates significantly above chance. This pattern may indicate \say{hot spots} for funders, where they converge on specific research areas. As illustrated in~\Cref{fig:topic_centers}, NSF has a 23\% funding rate for central topics related to multimodal analytics and outcome prediction, while institutional support appears to fund outcome prediction (25\%) and social learning (23\%) at a similar rate.

To assess whether the funding-topic network structure evolved, we calculated bipartite networks for each year 2020--2025. The funder-topic center space was relatively stable from 2020 to 2022, with a slight dip in connectivity between topics and funders. After 2022, there was a surge, peaking in 2024 with 26 edges, a density of 0.217, and an average degree of 3.25. Thus, in 2024, funding activity increased: funders supported more topics, and topics attracted support from a greater number of funders. In 2025, connectivity slightly decreased, returning to a level similar to 2023. 

The modularity was highest in 2021 (\textit{M} = 0.366), suggesting that the funding landscape was more compartmentalized. This might mean funders focused on specific, separate topics, or that topics were funded by a small number of specialized funders. The modularity reached its lowest point in 2024 (\textit{M} = 0.166), coinciding with the network's highest connectivity. This is a key finding: as the network became more connected, the distinct communities became less clear. This could be due to funders diversifying their portfolios across different topics, or topics attracting a broader range of funders, leading to a more integrated ecosystem where the lines between specialized areas are less rigid.

In summary, we found that the topic space of LAK research over the last six years consisted of 29 topic communities interconnected with each other (\Cref{fig:topic_communities}). While larger communities of topics exist, smaller ones emerged and may indicate more specialized, niche research topics, such as data-driven game design elements. Next, abstracts of papers formed six clusters of topic centers with a significant relationship to funders, suggesting that specific funders may be more interested in certain topic centers compared to others. The top funders of topic centers were the National Science Foundation (23\% funding rate for outcome prediction and multimodal analytics), followed closely by institutional support funds provided by internal grants (25\% funding rate for outcome prediction and 23\% for social learning).

\section{RQ3: What Are the Topical Trends in LA Research Globally and Locally?}
\label{sec:rq3}

This research question examined how the topical landscape evolved over time and across countries. Beyond descriptive counts, our goal was to assess if LA is converging on shared thematic centers or fragmenting into nationally distinct agendas. We connected this inquiry to prior syntheses of paradigms influencing LA~\cite{baker2021four,Lang-handbook-la2022}, asking whether historical distinctions (e.g., LA vs. EDM emphases) persist or attenuate in recent years. We operationalize \emph{global} trends as changes in the volume and relative prominence of topic centers across time (see \Cref{subsec:rq2-results} for a description of these centers), and \emph{local} trends as country-specific topical profiles and cross-country similarity. Convergence is indicated by (i) stable topical centroids over periods, (ii) increasing similarity between countries within the same topic, and (iii) stronger coupling between topical similarity and cross-country collaboration. Fragmentation would be reflected in shifting centroids, widening cross-country dispersion within topics, and weak alignment between similarity and collaboration.

\subsection{Methods}

Building on the six topic centers (clusters) from RQ2, we (i) track topic volumes and multi-country rates by period; (ii) quantify temporal stability via cross-period cosine similarity of topic centroids; (iii) quantify \emph{local} alignment via cross-country cosine similarity within each topic; and (iv) test whether topically similar countries collaborate more (Spearman correlation between country–country topical similarity and co-authorship counts). These indicators show whether the field exhibits thematic convergence globally while retaining (or not) local variation in fast-moving areas.

To examine the relation between topical similarity and international collaboration, we created a country–country collaboration matrix with the number of co-authored papers between each pair of countries. We also computed pairwise cosine similarities between country embedding vectors from the topical representation learning (see \Cref{subsec:rq2-representation-learning}). For each country pair, we combined the similarity score with the count of co-authored papers. Collaboration counts were log-transformed to account for skewness in their distribution. Then, we assessed the association between embedding similarity and collaboration using Spearman correlation, a non-parametric measure of monotonic association.

\subsection{Results}

We observed a strong, statistically significant positive correlation between embedding similarity and the extent of cross-country collaboration ($\rho = 0.44$, $p < 0.001$). Countries with more similar topical profiles are also more likely to co-author papers at LAK. The strongest collaborations align with very high topical similarity: Australia--UK (35 papers with a similarity of 0.98), followed by the USA--Canada (23 papers, 0.97), the USA--Australia (22 papers, 0.96), Australia--Brazil (17 papers, 0.87), and the USA--UK (15 papers, 0.94). Moreover, when countries co-author frequently, their topical profiles were highly similar (mean similarity for collaborating pairs $M = 0.84$), compared to pairs without collaborations ($M = 0.65$). The spread of similarities was wider for non-collaborating pairs (SD = 0.16), indicating that some highly similar countries (e.g., Switzerland with France or Israel, $>0.82$) have not yet co-authored at LAK.

\begin{figure*}[!ht]
\centering
\includegraphics[width=\linewidth,trim=0 0 0 20,clip]{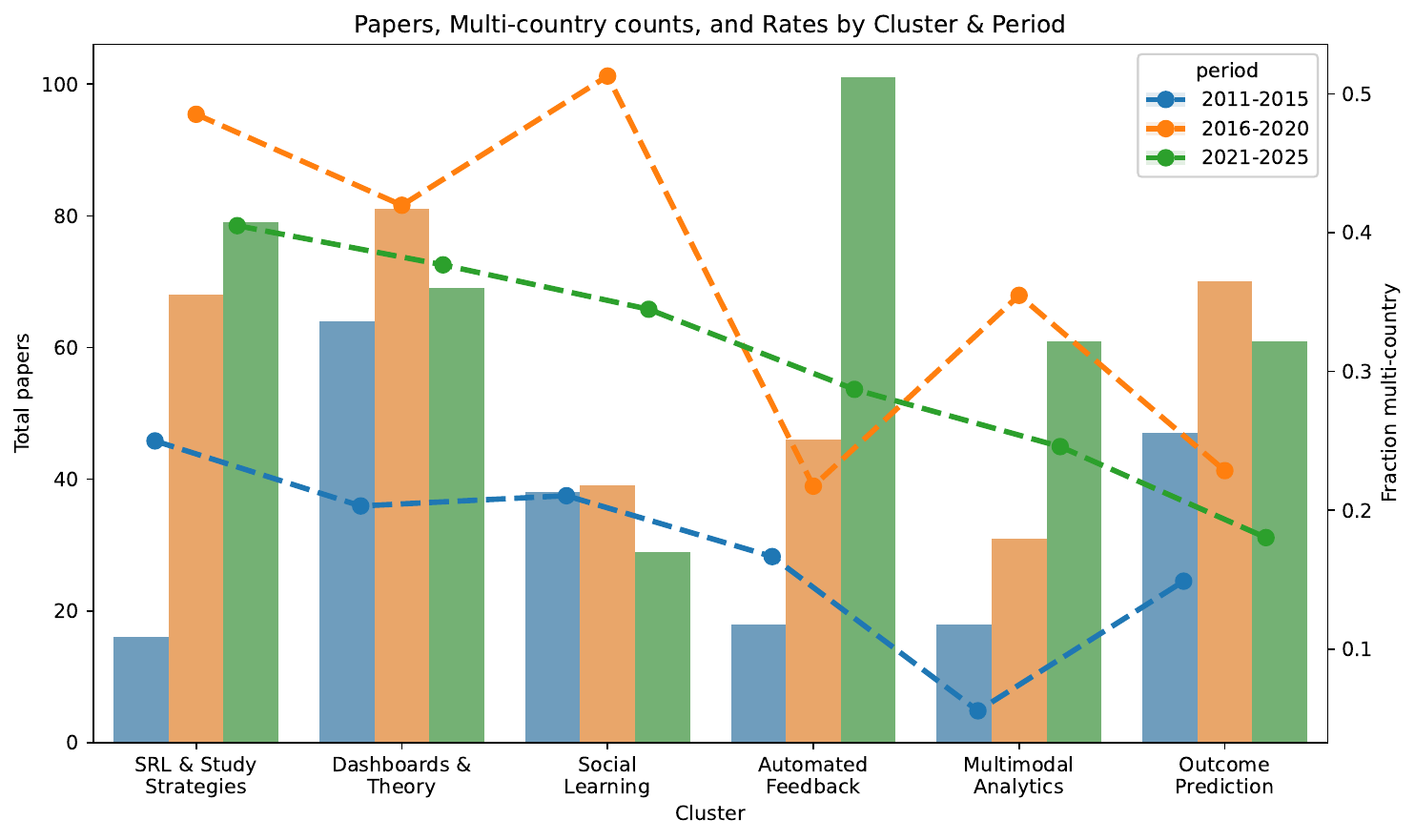}
\caption{Frequency (bar plots; left y-axis) and relative share of papers with more than one contributing country (line graphs; right y-axis) by period and identified topic cluster.}
\Description{For each of the six topic centers, their multi-country fractions are shown across the three 5-year periods.}
\label{fig:collaboration_trend_topics}
\end{figure*}

Regarding the topical centers' distribution (see \Cref{fig:collaboration_trend_topics}), the largest cluster was \textit{Dashboards \& Theory} (Cluster1; 214 papers), followed by \textit{Outcome Prediction} (Cluster5; 178), \textit{Automated Feedback} (Cluster3; 165), \textit{SRL \& Study Strategies} (Cluster0; 163), \textit{Multimodal Analytics} (Cluster4; 110), and \textit{Social Learning} (Cluster2; 106). Topics showed substantial variation in (a) when they were most published on and (b) how many countries collaborated on them per paper. Based on \Cref{fig:collaboration_trend_topics}, topical trajectories diverged across periods. \textit{SRL \& Study Strategies} (the least represented in 2011--2015) grew steadily and remains one of the largest clusters. This represents a turn toward stronger and more foundational guiding theory in LA research, which is considered important by the community~\cite{Wise_Shaffer_2015}. In contrast, both \textit{Dashboards \& Theory} and \textit{Outcome Prediction} expanded strongly through 2020 but have recently declined in relative prominence. While dashboards held initial promise of impacting learners, this has since been more strongly realized in feedback analytics. This was evidenced by \textit{Automated Feedback} surging in 2021–2025, emerging as the most prominent topical center in the field currently. Finally, \textit{Multimodal Analytics} has shown a stark increase in recent years, whereas \textit{Social Learning} peaked earlier (2016–2020) and has since tapered off due to declines in methods such as social network analysis. 

International collaboration also varied substantially by topic, based on the ratio of papers with more than one contributing country, called \say{fraction multi-country} (see \Cref{fig:collaboration_trend_topics}). \textit{Automated Feedback} has been increasingly multi-country collaborative in line with its surge in frequency. \textit{Outcome Prediction}, on the other hand, peaked in 2016–2020 before declining to one of the lowest collaboration rates by 2021–2025. \textit{SRL \& Study Strategies} shows relatively high and stable collaboration rates across the whole period, while \textit{Multimodal Analytics}—despite its recent growth in paper counts—remains characterized by comparatively low and declining levels of cross-country authorship. Finally, \textit{Social Learning} has seen collaboration decline since its 2016–2020 peak (where it was also the most collaborative topic). These trends highlight that while some topical centers of LAK research were briefly globalized, sustained multi-country collaboration remains uneven and in several cases has receded in recent years.

\begin{figure*}[!ht]
\centering
\includegraphics[width=\linewidth]{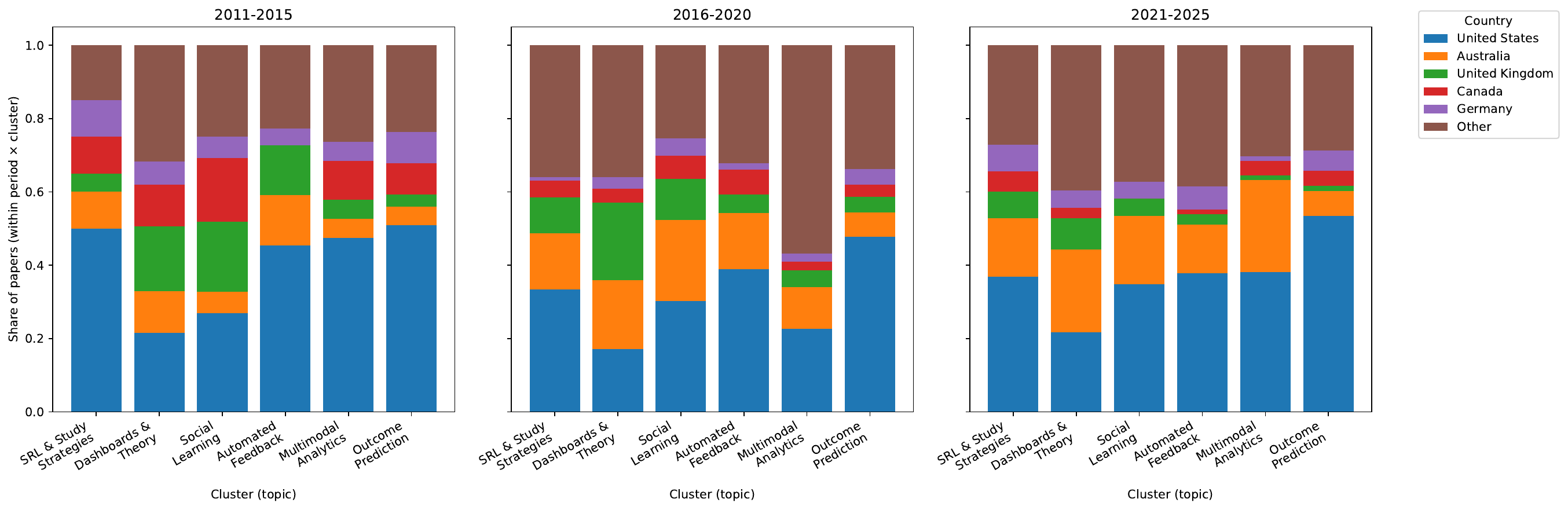}
\caption{Country contributions to topical clusters across three time periods (2011–2015, 2016–2020, 2021–2025). Bars represent the share of papers within each cluster, stacked by country.}
\Description{For each of the six topic centers, their country representation is shown across the three 5-year periods.}
\label{fig:country-content}
\end{figure*}

The topical composition of contributions varied markedly across countries (\Cref{fig:country-content}). The USA was consistently the largest contributor to \textit{Outcome Prediction}. Australia and the UK also played a prominent role, with Australia currently notably active in \textit{Dashboards \& Theory} and \textit{Multimodal Analytics}, and the UK maintaining a strong presence in \textit{Social Learning} and \textit{Dashboards \& Theory} (though with a declining trend). Canada contributed notably in 2011–2015, but its share has since diminished, while Germany has made modest but consistent contributions across most clusters (with a recent decline in multimodal analytics). The \say{Other} category, a mix of smaller or emerging contributors, has expanded over time. By 2021–2025, about one third of the output in many clusters—particularly \textit{Automated Feedback} and \textit{Dashboards \& Theory}—originates outside of the five top countries.  

The topical centroids are remarkably stable over time across clusters. Cosine similarities between periods are uniformly high, typically above 0.90. \textit{Dashboards \& Theory}, \textit{SRL \& Study Strategies}, and \textit{Social Learning} are the most stable, with cross-period similarities above 0.93–0.97 (with these three topics experiencing recent decline or small growth only). Clusters such as \textit{Automated Feedback}, \textit{Multimodal Analytics}, and \textit{Outcome Prediction} show slightly lower but still strong stability, with the largest early-to-recent shifts (around 0.90). These topics are marked by more recent growth (except outcome prediction). This shows that recently declining topics tended to change less.

The cross-country comparison of how similarly countries write about the same topical clusters indicates that clusters are widely shared across nations, though with some variation in alignment. For the largest clusters such as \textit{SRL \& Study Strategies} and \textit{Dashboards \& Theory}, pairwise similarities among the top five contributing countries are consistently very high ($>0.90$), with most values in the 0.92–0.98 range. This suggests that researchers in the US, Australia, the UK, Canada, and Germany frame these areas in highly consistent ways despite differences in the scale of their contribution. \textit{Social Learning} also shows strong convergence (typically 0.93–0.96), though Germany’s positioning falls slightly lower (around 0.86–0.89), indicating somewhat distinctive emphases.

Clusters that have recently surged — notably \textit{Automated Feedback} and \textit{Multimodal Analytics} — show greater dispersion. In \textit{Automated Feedback}, similarities between the USA, Australia, and the UK remain high (0.91–0.94). However, alignments with Germany and Brazil dip as low as 0.82, suggesting a diverse range of approaches and contexts. \textit{Multimodal Analytics} is even more fragmented, with USA–Australia similarities above 0.90 but pairings with Canada and the Netherlands as low as 0.76–0.81, reflecting distinctive national trajectories in multimodal data collection and analysis. \textit{Outcome Prediction} falls between these two extremes: USA, Australia, and the UK align above 0.90, but Canada and Germany show lower similarity (0.83–0.85), hinting at national differences in predictive modeling emphases or data sources. 

When examining abstracts of papers with contributors from underrepresented countries, we sometimes observed moderate similarity (0.53--0.59) to the top contributors. For example, one paper with Taiwanese authors in \textit{Automated Feedback} explores semantic indexing of paper-based exams. This context-specific solution diverges from the dominant focus on automated feedback for essays or open responses in larger-country work. Its lower similarity (0.53) reflects this framing while still aligning with the feedback cluster. Likewise, a paper with Moroccan authors in \textit{SRL \& Study Strategies} examines dropout behavior in a science learning game, linking gameplay errors and survival analysis to SRL processes. However, the emphasis on game-based learning related to SRL results in a distinct topical representation (similarity 0.54). These cases demonstrate how contributions from underrepresented contexts introduce distinctive methodological or application perspectives that enrich the field, even if they diverge from the mainstream.

\section{Discussion and Takeaways}

Building on prior reflective studies \cite{Kaliisa2024dashboardshype, Ferguson2025doctoral}, we examined trends across 936 LAK papers over a 15-year period. Our study provides methods and insights to strengthen LA research and practice, with implications for applications and learning.

\subsection{Summary of Key Results and Their Implications}
\label{subsec:implications}

Analyzing 15 years of LAK revealed a field that is both stable and permeable. A small core of prolific authors provides continuity, and a third of authors publish more than once; collaboration has expanded among authors and countries (RQ1). Topic modeling and keyword communities converge on six enduring centers: self-regulated learning, dashboards and theory, social learning, automated feedback, multimodal analytics, and outcome prediction. The funding–topic network showed systematic alignment between funders and centers, becoming more integrated by 2024 as modularity declined (RQ2). Topical centroids are highly stable over time; however, centers follow different trajectories and patterns of international co-authorship. Similarity in topical profiles is associated with higher collaboration (RQ3).

Our findings carry three implications. 
First, the overall stability of established centers contrasts with the divergence of contributions from underrepresented countries. The field can benefit from highlighting these differences, as they may reveal how LA practices adapt to local contexts and bring research closer to application, thereby closing the loop \cite{Motz2023LA, Clow2012cycle}. Neighboring fields, such as AIED, launched local initiatives that increase impact, such as low-tech \say{AIED Unplugged}~\cite{isotani2023aied} and dedicated conference tracks aimed at underrepresented contexts, such as \say{WideAIED} \cite{cristea2025artificial}. While Learning Analytics Summer Institutes and Learning Analytics in Practice are valuable for capability building and translation, dedicated LAK tracks or forums could make such contributions more visible and amplify what can be learned from them. 

Second, the evolving funding structure provides leverage for cross-center work. The increase in connectivity and decline in modularity in 2024 suggest funders support broader portfolios, which opens opportunities to design calls that bridge centers, for example, multimodal analytics that deliver actionable feedback to teachers. Such work could address longstanding critiques about limited impact on learning \cite{Kaliisa2024dashboardshype, Motz2023LA}. 

Third, we observe a rise in subfields with firm theoretical grounding, such as SRL, automated feedback, and multimodal analytics, alongside a decline in dashboards and outcome prediction after earlier hype cycles \cite{Kaliisa2024dashboardshype}. This shift signals progress toward theory-driven approaches, and also raises questions about how emerging technologies like LLMs may reshape the balance between support for metacognition and risks of over-reliance, including concerns about \say{metacognitive laziness} \cite{fan2025beware}. The field must consider how to sustain SRL as new tools change the learner’s role in the feedback process.

\subsection{Limitations and Future Work}

We analyzed only full and short LAK papers. This ensured comparable depth for our methods but excluded early-stage and applied work in companion venues, which could qualify the stability and convergence patterns (\Cref{sec:rq2,sec:rq3}). Topical trends in workshops could have predictive validity for forecasting future trends in the main proceedings. Next, funding acknowledgments incompletely capture institutional seed support and sometimes list only national programs, which could add noise to funder–topic links.
Future studies can add JLA papers and other LAK tracks to test if surges and declines (e.g., dashboards vs. automated feedback) occur outside the main track. 
Building on \Cref{sec:rq2}, targeted calls that bridge centers (e.g., multimodal analytics for actionable feedback) could be linked to in-the-wild evaluations aligned with the updated LA definition. 
Lastly, a promising direction is to develop a community dashboard to track topical, geographic, and funding trends. It could accelerate the translation into practice and be a living map of the field, as some scholars call for continuously updated \say{living literature reviews} to keep up with the changes \cite{wijkstra2021living}. 

\section{Conclusion}

This research mapped 15 years of LAK through an analysis of 936 papers, showing a consolidated and evolving field. Our results suggest three priorities. First, make underrepresented national contributions more visible to learn from locally grounded practice. Second, use the growing integration of funders to seed projects that bridge centers, for example, multimodal analytics that deliver actionable feedback. Third, pair theory-driven areas, such as SRL and automated feedback, with in-the-wild evaluation aligned to the actionable impact emphasized in the updated LA definition.

\begin{acks}
The work of Valdemar Švábenský on this paper was supported by the Czech Science Foundation (GAČR) grant no. 25-15839I. 
This research was partially supported by a Young Scholar grant from the Jacobs Foundation (project no: 2023151201) awarded to Elizabeth Cloude.
The research by Dragan Gašević was in part supported by the Australian Research Council (DP240100069 and DP220101209).
We thank Xavier Ochoa for compiling the LAK 2011--2019 dataset.
\end{acks}

\bibliographystyle{ACM-Reference-Format}
\makeatletter
\if@twocolumn
  \balance
\fi
\makeatother
\bibliography{references}

\end{document}